\documentclass[prd,showpacs,preprintnumbers]{revtex4} 
\usepackage{amsmath} 
\usepackage{amsfonts} 
\usepackage{amssymb}
\usepackage{graphicx} 
\usepackage[usenames]{color} 
\newcommand{\half}{\tfrac{1}{2}}
\usepackage{bbm}
\def\beq{\begin{equation}}
\def\eeq{\end{equation}}
\def\bea{\begin{eqnarray}}
\def\eea{\end{eqnarray}}
\def\ba{\begin{array}}
\def\ea{\end{array}}

\def\part{\partial}

\def\half{\frac{1}{2}}

\def\Re{{\mathfrak{Re}} }
\def\Im{{\mathfrak{Im}} }

\begin{document}

\preprint{UdeM-GPP-TH-08-165}
\preprint{arXiv: 0802.0354 [hep-th]}

\title{Path integration and perturbation theory with complex Euclidean actions}
\author{Garnik Alexanian$^1$}
\email{garnik@gmail.com}  
\author{R. MacKenzie$^2$}
\email{rbmack@lps.umontreal.ca}
\author{M. B. Paranjape$^{2,3}$} 
\email{paranj@lps.umontreal.ca}
\author{Jonathan Ruel$^4$} 
\email{ruel@fas.harvard.edu}

\affiliation{$^1$Troika Dialog, Romanov Pereulok 4, Moscow 125009, Russia}
\affiliation{$^2$Groupe de physique des particules, D\'epartement de physique,
Universit\'e de Montr\'eal,
C.P. 6128, Succ. Centre-ville, Montr\'eal, 
Qu\'ebec, CANADA H3C 3J7 }
\affiliation{$^3$Center for Quantum Spacetime, Department of
Physics, Sogang University, Shinsu-dong \#1, Mapo-gu,
Seoul, 121-742, Korea}
\affiliation{$^4$Department of Physics, Harvard University, \break
17 Oxford St., Cambridge, Massachusetts, USA, 02138}

\begin{abstract}  
The Euclidean path integral quite often involves an action that is not completely real {\it i.e.} a complex action.    This occurs when the Minkowski action contains $t$-odd  CP-violating terms. This usually consists of topological terms, such as the Chern-Simons term in odd dimensions, the Wess-Zumino term, the $\theta$ term or Chern character in 4-dimensional gauge theories, or other topological densities.   Analytic continuation to Euclidean time yields an imaginary term in the Euclidean action.  It also occurs when the action contains fermions, the fermion path integral being in general a sum over  positive and negative real numbers.   Negative numbers correspond to the exponential of $i\pi$ and hence indicate the presence of an imaginary term in the action.

In the presence of imaginary terms in the Euclidean action, the usual method of perturbative quantization can fail.  Here the action is expanded about its critical points, the quadratic part serving to define the Gaussian free theory and the higher order terms defining the perturbative interactions.  For a complex action, the critical points are generically obtained at complex field configurations.  Hence the contour of path integration does not pass through the critical points and the perturbative paradigm cannot be directly implemented.  The contour of path integration has to be deformed to pass through the complex critical point using a generalized method of steepest descent, in order to do so. Typically, this procedure is not followed.  Rather, only the real part of the Euclidean action is considered, and its critical points are used to define the perturbation theory, a procedure that can lead to incorrect results.

In this article we present a simple example to illustrate this point. The example consists of $N$ scalar fields in 0+1 dimensions interacting with a $U(1)$ gauge field in the presence of a Chern-Simons term. In this example the path integral can be done exactly, the procedure of deformation of the contour of path integration can be done explicitly and the standard method of only taking into account the real part of the action can be followed.  We show that the standard method does not give a correct perturbative expansion.  The implications of our work include perturbation theory in the standard model which is CP-violating, but also for calculations in the presence of topological terms which have given rise to radical changes in the spectrum of the theory. 
\end{abstract}

\maketitle

\section{Introduction\label{sec:intro}}

The Feynman path integral \cite{f} is a simple and powerful device for defining quantum field theory.  Its original definition, however, is mathematically delicate, to put it mildly.  The path integral is nominally defined as an integral expression
\beq
{\cal I}=\int{\cal D}\phi (t)\, e^{iS\left[\phi (t)\right]/\hbar},
\eeq
where $S\left[\phi (t)\right]$ is the action functional, which is completely real in Minkowski space, and it should give rise to the amplitude for a particular quantum mechanical process.  The function space over which we should integrate is usually infinite, but the integrand is always  unimodular, as long as the action is real. Therefore, the Feynman path integral cannot be absolutely convergent: 
\beq
\int{\cal D}\phi (t)\, \left| e^{iS\left[\phi (t)\right]/\hbar}\right|=\int{\cal D}\phi (t)\, 1=\nexists .
\eeq

It may be that the integral is convergent if it is performed over the integration domain in a specific order.  For example, consider a simple ordinary integral:
\beq
I=\int_{-\infty}^{+\infty} dx\,  e^{ix^2}\equiv\lim_{a\rightarrow\infty}\int_{-a}^{+a} dx\,  e^{ix^2}=\sqrt\pi .\label{integral}
\eeq
However, the limiting procedure treats preferentially the contribution from regions of the integration domain that are finite, relative to the contribution from those regions of the integration domain that are infinitely far away.  We are prescribed to integrate from $-a$ to $a$ first and then take the limit $a\rightarrow\infty$, although nothing intrinsic in the original integrand tells us to do so.  This prescription precludes the possibility of  the mathematical notion of integration as a measure defined on a given set, that maps functions on that set to the real numbers.  

If for example we choose to integrate first over the regions defined by 
\beq
S\equiv \left\{x \in \left[\sqrt {2 n\pi},\sqrt {2 n\pi+\frac{\pi}{4}}\right]\cup\left[-\sqrt {2 n\pi+\frac{\pi}{4}},-\sqrt {2 n\pi}\right]\quad, n=0,1,2,\cdots\right\},
\eeq
then we find divergent nonsense.  The exponential has positive real and imaginary parts  for each $n$.  Denoting $\Re (I_S)$ as  the real part of $I$ when the integration is restricted to $S$, since $\cos x^2\ge1/\sqrt 2$, we get
\beq
\Re(I_S)\ge 2\sum_{n=0}^{n=+\infty}\left(\frac{1}{\sqrt 2}\right)\left(\sqrt {2 n\pi+\frac{\pi}{4}}-\sqrt {2 n\pi}\right)=2\sum_{n=0}^{n=+\infty}\left(\frac{1}{\sqrt 2}\right)(\sqrt {2 n\pi})\left(\sqrt{1+\frac{1}{8n}}-1\right)=\infty \label{series}.
\eeq
This  is easy to see from the estimate
\beq
\sqrt{1+\frac{1}{8n}}>1+\half \frac{1}{8n}-\frac{1}{4}\left(\frac{1}{8n}\right)^2,\label{t}
\eeq
obtained from the fact that the Taylor expansion of $\sqrt{1+1/8n}$ around $n=\infty$, is an alternating (although absolutely) convergent series.  The second term from Eq.~(\ref{t}) is divergent when inserted in the final series of Eq.~(\ref{series}).

Another way of seeing the problem is to change variables $u=x^2$ in the integral of Eq.~(\ref{integral}), which yields
\beq
I=\int_{0}^{\infty} du\,  \frac{e^{iu}}{\sqrt u}.
\eeq
At the origin, there is an integrable singularity that poses no problem; however, for large $u$ the integral is equivalent to an alternating series whose terms are not absolutely convergent.   If we integrate successively from $u=n\pi/2$ to $u=(n+1)\pi/ 2$ with $n\in \mathbb{Z}$, the real and imaginary parts of the integral correspond to the sum of a decaying alternating series with terms that are not absolutely convergent.  Looking at the real part, for example, gives the alternating series
\beq
\Re (I)=\int_{0}^{{\pi/2}}\hskip-.5cm du\,  \frac{|\cos u|}{\sqrt u} 
+\sum_{k=1}^\infty (-1)^k\int_{(2k-1){\pi/2}}^{(2k+1){\pi/2}}\hskip-1.3cm du \hskip1cm  \frac{|\cos u|}{\sqrt u} ,
\eeq
which is not absolutely convergent.  It is well known that such a series can be made to sum up to any number that one desires, simply by rearranging the order of the summation. 

Thus, the fundamental definition of the path integral as an integral over an infinite function space of an oscillating unimodular phase does not make mathematical sense.  In the limited examples where such an integral can actually be done, it is necessary to treat some parts of function space preferentially to other parts, by summing up their contribution first.  The function space, however,  should only be discriminated by the action.  The action should  govern which part of function space is physically important to a particular quantum amplitude.  Hence the Feynman path integral requires a more rigorous and meticulous definition.  In fact this is exactly what is done by analytical continuation to Euclidean time.  For more background and explanation see, for example,  the books by Glimm and Jaffe \cite{gj} or Simon \cite {s} and references therein.

We will consider mostly bosonic theories in this article.  Bosonic fields can always be decomposed into their real and imaginary parts and re-expressed entirely in terms of real fields.  When we treat fermionic theories, we will only consider the cases where the fermionic degrees of freedom can be also be re-expressed via bosonization \cite{c} or through the use of Chern-Simons \cite{djt} type actions in terms of bosonic fields. Again through decomposition into real and imaginary parts, we will always be able to write the action in terms of purely real bosonic fields. 

To define quantum mechanical amplitudes in  Minkowski space, we must use analytic continuation from Euclidean time.  Paradoxically, we start with the Minkowski action and analytically continue it to Euclidean time.   For actions made up of real bosonic fields the continuation to Euclidean time yields 
\beq
t\rightarrow -i\tau ,\quad \phi(t)\rightarrow\phi(\tau ),\quad 
{\cal I}\rightarrow\int{\cal D}\phi (\tau )\, e^{-S_E\left[\phi(\tau )\right]/\hbar}.
\eeq
Usually the Euclidean action $S_E\left[\phi(\tau )\right]$ is a real and positive definite functional of the Euclidean time field configuration.  However, for certain cases described in more detail in this paper,  the Euclidean action is not completely real and contains terms that are explicitly imaginary.  Such a situation, however, is not necessarily a fatal impediment to the definition of the functional integration.  If the real part of the Euclidean action can be used to define a measure on the space of field configurations, then the fluctuating part can be viewed as a bounded functional that can be integrated against this measure.  There are many examples where the path integral in Euclidean space can be rigorously proven to exist as an integration with respect to a measure on function space.  The corresponding Euclidean time amplitudes for relativistic theories satisfy a set of axioms \cite{w} that allow for these amplitudes to define properly a sensible quantum field theory.  We obtain the Minkowsi space amplitudes, or more generally the Minkowski space path integral, via analytically continuing the Euclidean space path integral and the corresponding amplitudes back to real time.  

However, in practice, even the Euclidean functional integration is most often evaluated in perturbation theory.  The Gaussian fluctuations about the critical points of the action are quantized without approximation, and the non-linear terms are treated as perturbations.  It is this procedure that can fail when the Euclidean action is not real.  The critical points of a complex action are in general at complex field values.  Thus the critical points are not attained for any real field configurations, and the quantization about these critical points cannot be addressed in terms of the strictly real fields with which we started. 

In such cases, there are two ways to proceed.  The first way is to consider only the real part of the action and find its critical points, which  are obtained at real field configurations.  Then the fluctuations about these  critical points can be quantized, leaving the imaginary part of the action and the non-linear terms to give rise to the perturbation theory.  The second  way is to analytically continue the ``contour" of functional integration into the space of complex field configurations so that it passes through the complex critical point and to use the contour of path integration that corresponds to the ``contour of steepest descent".  Then the fluctuations of these complex field configurations about the complex critical point are quantized, leaving only the non-linear terms to define the perturbation theory.  Although this idea has already been considered in  \cite{tsh},and in \cite{kml}, but it has not been fully implemented in any concrete example. 

The main results of the present paper were announced in \cite{ampr}.  Here we expand on the theme and show all the details.   The model we consider,  0+1-dimensional ``electrodynamics" with $N$  scalar fields and a Chern-Simons term,  can be solved exactly, and it can also be treated using both of the perturbation schemes outlined above.  We find that  the first procedure (expanding about the critical points of the real part of the action) does not work, while  the second procedure (expanding about the critical points of the full, complex action) does in fact give the correct perturbative expansion.  

\section{Complex Euclidean actions\label{2}}
Complex Euclidean actions may arise for many reasons but we will only consider the case where they occur due to a $t$-odd term in the Minkowski action.  Such a term, under analytic continuation to Euclidean time, will necessarily be imaginary.   Then the action will split into two parts, 
\beq
S\left[\phi (t)\right]= S_{even}\left[\phi (t)\right]+S_{odd}\left[\phi (t)\right],
\eeq
where $S_{even}\left[\phi (t)\right]\rightarrow S_{even}\left[\phi (t)\right]$ but $S_{odd}\left[\phi (t)\right]\rightarrow -S_{odd}\left[\phi (t)\right]$ for $\phi (t)\rightarrow \phi (-t)$.  $S_{odd}\left[\phi (t)\right]$ may take the guise of
\beq
S_{odd}\left[\phi (t)\right]\sim
\begin{cases}
\int dt d^2x\,\epsilon^{\mu\nu\lambda}(A_\mu\partial_\nu A_\lambda +\cdots)\cr 
\int_{half\,S^5} dtd^4x\,\epsilon^{\mu\nu\lambda\sigma\tau}Tr (U^\dagger\partial_\mu U\cdots U^\dagger\partial_\tau U )\cr 
\int dt d^3x\,\epsilon^{\mu\nu\lambda\sigma}Tr( F_{\mu\nu}F_{\lambda\sigma})
\end{cases}
\eeq
corresponding to a Chern-Simons term in 2+1 dimensions \cite{djt}, a Wess-Zumino-Novikov-Witten (WZNW) term in 3+1 dimensions \cite{wz},\cite{n},\cite{wi}, and the $\theta$ term in 3+1 dimensions respectively give rise to imaginary terms in the Euclidean action.  All of these terms contain only one time derivative; the $t$-odd part of the action has the form
\beq
\int dt\int d^dx \cdots\partial_t\cdots
\eeq
in $d+1$ dimensions.  On analytic continuation to Euclidean time, $t\rightarrow -i\tau$ but $\partial_t\rightarrow i\partial_\tau$; thus, the odd part of the action is invariant 
\beq
\int dt\int d^dx \cdots \partial_t\cdots\rightarrow\int d\tau\int d^dx \cdots\partial_\tau\cdots
\eeq
so that
\beq
\int dt\int d^dx L_{odd}^{Minkowski}\left[\phi (t)\right]\rightarrow \int d\tau\int d^dx L_{odd}^{Euclidean}\left[\phi (\tau)\right],
\eeq
while the even part of the action satisfies
\beq
\int dt\int d^dx  L_{even}\left[\phi (t)\right]\rightarrow i\int d\tau\int d^dx  L_{even}\left[\phi (\tau)\right],
\eeq
where $L_{even}\left[\phi (\tau)\right]$ is by hypothesis a real, positive functional, which means that an overall minus sign has been factored out after the analytic continuation.  Then the Euclidean functional integral becomes
\beq
{\cal I}=\int{\cal D}\phi (\tau )\, e^{-(\int d\tau\int d^dx L_{even}\left[\phi(\tau )\right]-i\int d^dx L_{odd}\left[\phi(\tau )\right])/\hbar}.
\eeq
This Euclidean functional integral is in principle well defined, in that the real part of the action serves to define a measure on the space of field configurations.  However the perturbative quantization is not straightforward as the critical points of the action are in general achieved at complex values of the field configurations.  Below we flesh out the details of two examples where the Euclidean action is complex and critical points are generally obtained for complex values of the fields.   

\subsection{Abelian Maxwell-Chern-Simons-Higgs theory}
In 2+1 dimensions, the abelian Maxwell-Higgs theory, scalar electrodynamics, admits the interesting possibility of adding the Chern-Simons \cite{djt} term to the Lagrangian.  This term is a topological term: it does not actually depend on the metric of space-time.  It is, however, not a total derivative; hence, it contributes to the equations of motion.  The interesting feature of this term is that it allows for a mass term for the gauge field in a gauge-invariant manner, and it allows for fractional statistics for the charged particles.  The scalar field no longer represents bosons, but anyons \cite{wil}, which have an arbitrary change of phase that is fixed by the coefficient of the Chern-Simons term under the exchange of two such particles.  For an appropriate value of the coefficient of the Chern-Simons term, the scalar field actually describes fermions.  Hence, the path integral here is an explicit example in which fermions are considered, and the question of the sign problem of the fermion determinant is treated. 
The Minkowski action is given by using two real scalar fields to represent the complex scalar field
\beq
S_{Mink.}=\int dtd^2x \frac{1}{2}(\partial_\mu \delta_{ij}-eA_\mu\epsilon_{ij})\varphi_j(\partial^\mu \delta_{ik}-eA^\mu\epsilon_{ik})\varphi_k -\lambda(\varphi_j\varphi_j-a^2)^2 -\frac{1}{4}F_{\mu\nu}F^{\mu\nu} +\theta\epsilon^{\mu\nu\lambda}A_\mu\partial_\nu A_\lambda,
\eeq
where $F_{\mu\nu}=\partial_\mu A_\nu -\partial_\nu A_\mu$ and $\epsilon_{ij}$ and $\epsilon^{\mu\nu\lambda}$ are the Levi-Civita symbols for two and three indices respectively.  The analytic continuation to Euclidean space proceeds with the replacement $t\rightarrow -i\tau$ which generates $\partial_t\rightarrow i\partial_\tau$, and for consistency we must also transform $A_0(t)\rightarrow iA_3(\tau )$ (note we are in 2+1 dimensions).  This yields the transformation
\bea
iI_{Mink.}&\rightarrow &i (-i)(-)\int dtd^2x \left(\frac{1}{2}(\partial_\mu \delta_{ij}-eA_\mu\epsilon_{ij})\varphi_j(\partial_\mu \delta_{ik}-eA_\mu\epsilon_{ik})\varphi_k +\lambda(\varphi_j\varphi_j-a^2)^2 \right.\cr
 &+&\left.\frac{1}{4}F_{\mu\nu}F_{\mu\nu} - i\theta\epsilon_{\mu\nu\lambda}A_\mu\partial_\nu A_\lambda\right)\cr
 &\equiv& -I_{Euclidean}.
\eea
Of note is the explicit appearance of the imaginary number $i$ appearing in the Euclidean action, which is otherwise completely real and positive.  The equations of motion are
\begin{eqnarray}
(\partial_\mu \delta_{ij}-eA_\mu\epsilon_{ij})(\partial_\mu \delta_{jk}-eA_\mu\epsilon_{jk})\varphi_k +4\lambda (\varphi_j\varphi_j -a^2)\varphi_i&=&0,\cr
\partial_\nu F_{\mu\nu} -i\theta\epsilon_{\mu\nu\lambda}\partial_\nu A_\lambda + e\epsilon_{ik}\varphi_k(\partial_\mu-eA_\mu\epsilon_{ij}\varphi_j),
&=&0 \end{eqnarray}
which contain an explicit imaginary term.  Amp\`ere's law and Gauss' law for static field configurations give
\bea
\partial_mF_{lm}- i\theta\epsilon_{li0}\partial_i A_0+e\epsilon_{ik}\varphi_k(\partial_l-eA_l\epsilon_{ik})\varphi_k&=&0,\cr
-\partial_i\partial_i A_0 -i\theta\epsilon_{0ij}\partial_i A_j + e^2A_0\varphi_j\varphi_j&=&0,
\eea
which has an obvious solution with imaginary $A_0$ and real $A_i$.  The generalization of the vortex solution in this theory necessarily requires imaginary $A_0$.  

\subsection{Non-abelian Goldstone bosons and the Wess-Zumino term}
Another example of complex action is given by the 1+1-dimensional model of non-abelian Goldstone bosons \cite{wi2}, a non-linear sigma model with the WZNW term added.  Here the field variable is a group-valued unitary matrix, $g(x,t)$ and the Minkowski action is
\beq
I_{Mink.}=\frac{1}{8\pi}\int dtdx\, Tr\left[\partial_\mu g^\dagger\partial^\mu g\right] +\frac{N}{12\pi}\int dtd^2x\, \epsilon^{JKL}Tr\left[ g^\dagger\partial_J gg^\dagger\partial_K gg^\dagger\partial_L g\right],\label{wz}
\eeq
where the first integral is over 1+1-dimensional space-time while the second one is over half a three-sphere whose boundary is 1+1-dimensional space-time. The indices $\mu , \nu ,\cdots $ etc. take the values $0,1$ while the indices $J,K,\cdots$ take the values $0,1,2$, and $N$ is an integer.  The integrand of the second term is a total derivative; hence, the integrand essentially depends only on the  1+1-dimensional boundary values of the field.  Its variation under compact infinitesimal deformations of the field variables is an explicit total derivative, hence its contribution to the equations of motion is a 1+1-dimensional local term.  However the actual value of the integral for fixed boundary values is ambiguous.  It depends on how we extend the boundary configuration into the half-three-sphere and can change by an integer times a fixed value.  Hence the quantum mechanical path-integral only makes sense if the coefficient of the term is appropriately quantized so that the change in the action is:
\beq
I_{Mink.}\rightarrow I_{Mink.} +2\pi k,
\eeq
where $k$ is an integer, so that $e^{iI_{Mink.}}$ is invariant.  Choosing the normalization as we have in Eq.~(\ref{wz}) ensures this transformation property.  It is evident then, that the coefficient of the WZNW term in Euclidean space must be imaginary if the Euclidean path integral is to make any sense.  Indeed, since it is $t$-odd, it remains imaginary on analytic continuation to Euclidean space, 
\beq
I_{Eucl.}=\frac{1}{8\pi}\int d\tau dx\, Tr\left[\partial_\mu g^\dagger\partial_\mu g\right] -i\frac{N}{12\pi}\int d\tau d^2x \,\epsilon_{JKL}Tr\left[ g^\dagger\partial_J gg^\dagger\partial_K gg^\dagger\partial_L g\right] ,
\eeq
where now, $\mu , \nu ,\cdots $ etc. take the values $1,2$ while the indices $J,K,\cdots$ take the values $1,2,3$.  The Minkowski equations of motion for $N=1$ are
\beq
 \frac{2}{8\pi}\partial^\mu \left[g^\dagger\partial_\mu g\right]+ \frac{3}{12\pi} \epsilon^{\mu\nu}\left[ g^\dagger\partial_\mu gg^\dagger\partial_\nu g\right] =0,
 \eeq
which can be written in the compact form
 \beq
( \partial^\mu +\epsilon^{\mu\nu}\partial_\nu )\left[g^\dagger\partial_\mu g\right] =0,
 \eeq
or, more explicitly,
 \beq
 (\partial_0+\partial_1)\left[g^\dagger  (\partial_0-\partial_1)g\right]\equiv \partial_+\left[g^\dagger  \partial_-g\right]=0.
 \eeq
 The general solution is $g=A(x_+)B(x_-)$ for two arbitrary unitary matrix valued functions $A(x_+)$ and $B(x_-)$ where $x_\pm=x^0\pm x^1$.  The Euclidean equations, however,  become:
 \beq
( \partial_\mu -i\epsilon_{\mu\nu}\partial_\nu )\left[g^\dagger\partial_\mu g\right] =0\label{ee}
 \eeq
 which simplify to
 \beq
( \partial_1 -i\partial_2 )\left[g^\dagger( \partial_1 +i\partial_2 ) g\right]\equiv  \partial_z\left[g^\dagger  \partial_{\bar z} g\right]=0
 \eeq
 where $z=x^1+ix^2$ and $\partial_z =\partial_1 -i\partial_2 $.  The putative solution, $g=A( z)B(\bar z)$ where $A( z)$ is a unitary matrix valued holomorphic function and $B(\bar z)$ is a unitary matrix valued anti-holomorphic function, simply does not work since no such functions exist, except for trivial constant functions.  Non-trivial solutions exist if we allow for extension into complex valued field configurations.  In this case, we must allow $A(z)$ to be an arbitrary holomorphic function with values in $GL(n,\mathbb{C})$, the general linear group of some fixed dimension $n$ over the field of complex numbers,  while $B(\bar z)$ is correspondingly an arbitrary anti-holomorphic function.  This equivalently expresses the fact that Eq.~(\ref{ee}) has no solution for $g^\dagger\partial_\mu g$ in the (anti-hermitian) Lie algebra of the the unitary group $\mathfrak{u}(n)$, one term is hermitian but the other term is anti-hermitian in this equation. The equation does admit solutions in its complexified extension to the Lie algebra of the general linear group, $\mathfrak{gl}(n)$.  The necessity to complexify the space of field configurations was observed by Di Vecchia et al \cite{d} but the functional integral was not addressed.
 
 In the next section we treat a simple model where we can perform the functional integral both exactly and in Gaussian approximation about the complex critical points and we can see how it is necessary to analytically continue the contour of functional integration to complex field configurations in order to evaluate the integral perturbatively.

\section{0+1-dimensional Abelian Higgs model with Chern-Simons term\label{3}}
In this section we treat a model which encompasses all the aspects of the problem of Euclidean complex actions, with the important simplicity that the model can be exactly solved, as well as solved using various perturbation schemes.  This allows us to test directly whether it is correct to define perturbation theory by expanding only about the critical points of the real part of the Euclidean action or do we have to expand about the critical points of the full complex Euclidean action, thereby encountering, in general,  complex critical points.  We emphatically find that it is incorrect to define the perturbation theory by expanding about the critical points of only the real part of the Euclidean action.  This result could, in principle, have important consequences for perturbative calculations of the phenomenology in $CP$-violating theories, such as the standard model.  

In 0+1 dimensions, quantum field theory is actually equivalent to quantum mechanics. We will nevertheless treat this model as a quantum field theory and continue the time $t$ to Euclidean time $t\rightarrow -i\tau$.  We consider $N$ massive, charged scalar fields $\phi_i$ which interact with an abelian gauge field $A$, both of which are functions of the Euclidean time $\tau$.   The gauge field kinetic term  is absent, as we cannot define $F_{\mu\nu}$ in $0+1$ dimensions.  However we can add a Chern-Simons term, which is simply proportional to the gauge field $A$.  The Euclidean action density is given by
\beq
{\cal L}=\sum_{i=1}^N\left(\left| (\partial_\tau +iA)\phi_i\right|^2+m^2\left|\phi_i\right|^2\right) +i\lambda A.
\eeq
where $\lambda$ is just a parameter.  We take compact Euclidean time, 
\beq
\tau\, \colon\, 0\rightarrow\beta
\eeq
which corresponds to finite temperature $\beta ={1\over k_B T}$.  We take the analog of Lorentz gauge, 
\beq
\partial_\tau A=0\, \Rightarrow A\equiv constant.
\eeq
Then topologically nontrivial gauge transformations $U=e^{i2\pi n\tau/\beta}$, where $n$ is an integer have the effect of shifting $A\rightarrow A-2\pi n/\beta$.  This means that we can restrict $A\in 0\rightarrow 2\pi/\beta$.  The Chern-Simons term, however, is not invariant:
\beq
i\lambda\int_0^\beta d\tau A\rightarrow i\lambda\int_0^\beta d\tau (A-\partial_\tau\Lambda )=i\lambda\int_0^\beta d\tau (A-2\pi n/\beta )=i\lambda(\beta A-2\pi n ).
\eeq
The exponential of the Chern-Simons term however, can be made invariant if the coefficient $\lambda$ is quantized to be an integer, $\lambda =N$.  We choose the same integer as the number of scalar fields; it is clear, however, that we could have chosen a different integer for the Chern-Simons term.   We will be considering the limit $N\rightarrow\infty$ as it is in that limit that we can do the steepest descent (Gaussian) approximations that define perturbation theory.  If the integers chosen are different, then they both must simultaneously become large.  For the sake of clarity, we will not consider this detail here. 

The critical points of the action are attained at solutions of the equations of motion:
\beq
-D_\tau^2\phi_i +m^2\phi_i =0\quad\quad
\int_0^\beta d\tau \sum_{i=1}^N\left( i(\partial_\tau\bar\phi_i\phi_i -\bar\phi_i\partial_\tau\phi_i )+2A\bar\phi_i\phi_i\right) -iN\beta =0.
\eeq
The second equation can easily be solved and exhibits the complex critical point; the solution $A^{*}$ is given by
\beq
A^{*}={iN\beta -  \int_0^\beta d\tau  i\sum_{i=1}^N(\partial_\tau\bar\phi_i\phi_i -\bar\phi_i\partial_\tau\phi_i )\over 2\int_0^\beta d\tau \sum_{i=1}^N\bar\phi_i\phi_i}\approx {i\gamma +\alpha\over \delta}
\eeq
for some real values of $\gamma$, $\alpha$ and $\delta$. (Note that the term $i\sum_{i=1}^N(\partial_\tau\bar\phi_i\phi_i -\bar\phi_i\partial_\tau\phi_i )$ is real.)  The scalar field equations actually have no solution on the space of periodic functions for a gauge field of nontrivial holonomy, $A\ne 2n\pi /\beta$.  This 
fact is not a problem, as we can simply perform the scalar field functional integrals exactly for any value of the gauge field, and hence eliminate them.  Each of the scalar fields can be expressed in terms of Fourier modes,
\beq
\phi (\tau )=\sum_{n=-\infty}^\infty\varphi_ne^{i2\pi n\tau /\beta },
\eeq
giving
\beq
\int_0^\beta d\tau\left| D_\tau\phi\right|^2 =\beta\sum_{n=-\infty}^\infty\bar\varphi_n\varphi_n ((2\pi n /\beta )+A)^2.
\eeq
This yields each scalar field functional integral as
\beq
Z_1(\beta ,m,A)=\int_{-\infty}^\infty\prod_n d\{ \bar\varphi_n\varphi_n\}e^{-\beta\sum_n\left|\varphi_n\right|^2 (({2\pi n \over\beta} +A)^2+m^2)}.
\eeq
Integrating over the Gaussian fluctuations of each Fourier mode gives the infinite product
\beq
Z_1(\beta ,m,A)=\prod_{n=-\infty}^\infty {\pi\over\beta (({2\pi n \over\beta} +A)^2+m^2)}. 
\eeq
The product evidently vanishes, but when normalized with respect to $A=0$ we obtain
\beq
{Z_1(\beta ,m,A)\over Z_1(\beta ,m,0)}=\prod_{n=-\infty}^\infty {({2\pi n \over\beta} )^2+m^2\over ({2\pi n \over\beta} +A)^2+m^2}
\eeq
which is a finite, well-defined product.  It is well known how to evaluate the product; adapting methods found in  \cite{j} and \cite{dll}  gives us, for $N$ scalar fields,
\beq
{Z(\beta ,m,A)\over Z(\beta ,m,0)}=\left({\cosh\beta m -1\over \cosh\beta m -\cos\beta A}\right)^N.
\eeq
Thus the ``functional" integral that we are left with is
\beq
{\cal I}(N,\beta ,m)=\int_0^{2\pi /\beta }dA\left({\cosh\beta m -1\over \cosh\beta m -\cos\beta A}\right)^N e^{iN\beta A}.\label{fi}
\eeq
This integral can be performed exactly, but also in the limit $N\rightarrow\infty$ in a saddle point approximation.  The critical points of the action are achieved at complex values of the integration variable $A$.  Therefore we can compare the exact  functional integral with a saddle point approximation using the full action and with a saddle point approximation using only the real part of the action to find the critical points. We will find that using only  the real part of the action gives erroneous results.

\subsection{Exact result}
To perform the integral exactly, we use the complex variable $z=e^{i\beta A}$.  Then $dA=dz/i\beta z$ and the integral becomes a contour integral over the unit circle in the complex $z$ plane:
\beq
{\cal I}(N,\beta ,m)=\oint {dz\over i\beta}\left({2(\cosh\beta m -1)\over 2z\cosh\beta m -z^2-1}\right)^N z^{2N-1}.
\eeq
The poles of the integrand simplify nicely; they are
\beq
z_\pm^*=\cosh\beta m\pm\sqrt{(\cosh\beta m)^2 -1}=e^{\pm\beta m}.
\eeq 
$z_+^*$ is outside the unit circle while $z_-^*$ is inside; thus, only the latter contributes.  The residue is calculated by the formula
\beq
{\cal I}(N,\beta ,m)={2\pi(\cosh\beta m -1)^N2^N\over \beta (-1)^N(N-1)!}\sum_{k=0}^{N-1}\left({N-1\atop k}\right)\left. \left({d^kz^{2N-1}\over dz^k}\right)\left({d^{N-1-k}\over dz^{N-1-k}}{1\over (z-e^{\beta m})^N}\right)\right|_{e^{-\beta m}}.
\eeq
This expression can be worked out further to give the exact result
\beq
{\cal I}(N,\beta ,m)={2\pi(\cosh\beta m -1)^N2^N\over \beta (N-1)!}\sum_{k=0}^{N-1}\left({N-1\atop k}\right) \left(\frac{(2N-1)!}{(2N-k-1)(N-1)!}\right)
\left({e^{-\beta m}\over e^{\beta m}-e^{-\beta m}}\right)^{2N-1-k}\label{ex}.
\eeq
In the limit $\beta m, N\rightarrow\infty$, we keep only the term $k=N-1$;  using the slightly extended Stirling approximation for the factorial, $N!\approx \sqrt{2\pi\over N}N^Ne^{-N}\left(1+o\left( \frac{1}{N}\right)\right)$, gives
\beq
{\cal I}(N,\beta ,m)\approx{e^{-N\beta m}2^{2N}\over \beta}\sqrt{\pi\over N}.\label{1}
\eeq

\subsection{Complex saddle point: critical point of the full action}
This result is reproduced perfectly by calculating the integral via the approximation of steepest descent.  Here the original integral from Eq.~(\ref{fi}), is written as
\beq
{\cal I}(N,\beta ,m)=(\cosh\beta m -1)^N\int_0^{2\pi /\beta }dA\,\, e^{-N\ln(\cosh\beta m -\cos\beta A)} e^{iN\beta A} =(\cosh\beta m -1)^N \int_0^{2\pi /\beta }dA\,\, e^{Nf(A)}
\eeq
where
\beq
f(A)=  -\ln (\cosh\beta m -\cos\beta A)+i\beta A  .
\eeq
The critical points are determined by imposing
\beq
\left.{\partial f(A)\over \partial A}\right|_{A^*}=f'(A^*)= {-\beta\sin\beta A^*\over \cosh\beta m -\cos\beta A^*} +i\beta=0\label{cp1}
\eeq
i.e.
\beq
e^{-i\beta A^*}=\cosh\beta m\label{cp2}
\eeq
which yields the complex critical points
\beq
\beta A^*=i\ln (\cosh\beta m) +2\pi k\label{cp3}
\eeq
where $k\in\mathbb{Z}$.   The original contour along the real axis from the origin, $A=0$ to $A={2\pi/\beta}$ is then to be deformed to the contour that first rises vertically from the origin $A=0$ along the imaginary axis to the critical point $A^*=i\ln (\cosh\beta m)/\beta$, then turning by 90$^\circ$ to follow the horizontal contour at $\Im A =\ln (\cosh\beta m)/\beta$ to $A^*=i\ln (\cosh\beta m)/\beta +2\pi/\beta $, and then turning again by 90$^\circ$ to descend along the line $\Re A =2\pi/\beta$ until $A=2\pi/\beta$.  The two vertical parts of the contour cancel each other while the horizontal part of the contour exactly leaves the critical point at $A^*=i\ln (\cosh\beta m)/\beta$ (arrives at $A^*=i\ln (\cosh\beta m)/\beta +2\pi/\beta $) along the path of steepest descent (ascent).  The contribution from each of these critical points is over half the Gaussian peak but  the two contributions simply combine to give the equivalent of integrating across just one of the critical points along the path of steepest descent.  The actual full contour of steepest descent continues into the complex plane in a rather complicated trajectory, however, in the saddle point approximation it is only the neighbourhood of the critical point that is relevant, where it leaves each critical point horizontally.  The method of steepest descent then gives the answer:
\beq
{\cal I}(N,\beta ,m)={(\cosh\beta m -1)^N}{e^{N(f(A^*)}}\sqrt{2\pi\over N\left|f^{\prime\prime}(A^*)\right|}.
\eeq
Hence  we will require the value at the critical point of the function, 
\bea\nonumber
e^{Nf(A^*)}&=&{e^{iN\beta A^*}\over (\cosh\beta m-\cos\beta A^*)^N}\\\nonumber
&=&{1\over (-i\sin(\beta A^*)^N\cosh^N(\beta m)}\\\nonumber
&=&\left({2\cosh\beta m\over \sinh^2\beta m}\right)^N{1\over \cosh^N\beta m}\\
&=&{2^N\over \sinh^{2N}\beta m},
\eea
where we have used Eqs. (\ref{cp1}-\ref{cp3}), and the second derivative at the critical point
\beq
f''(A^*)=-\beta^2\left({\cos\beta A^*\over \cosh\beta m-\cos \beta A^*}-{\sin^2\beta A^*\over (\cosh\beta m-\cos\beta A^*)^2}\right).
\eeq
Using Eqs. (\ref{cp1}-\ref{cp3}) again gives
\bea\nonumber
f''(A^*)&=&-\beta^2\left( {i\cos\beta A^*\over\sin\beta A^*}+1\right)\\
&=&-2\beta^2\coth^2\beta m\label{fpp1}.
\eea
Putting these together gives
\beq
{\cal I}(N,\beta ,m)\approx {(\cosh\beta m -1)^N}{2^N\over \sinh^{2N}\beta m}\sqrt{2\pi\over N\left( 2\beta^2\coth^2\beta m\right)}.
\eeq
This expression should be compared with the large $N$ limit of Eq.~(\ref{ex}), however the limit $\beta m\rightarrow\infty$ can be taken, as the Gaussian approximation  controlled by Eq.~(\ref{fpp1}) converges to zero.  This gives
\bea\nonumber
{\cal I}(N,\beta ,m)&\approx&\left({e^{\beta m}\over 2} \right)^N{2^N\over \beta({ e^{\beta m}/ 2})^{2N} }\sqrt{\pi\over N}\\
&=&{e^{-N\beta m}2^{2N}\over \beta}\sqrt{\pi\over N}
\eea
exactly the result obtained in Eq.~(\ref{1}).

\subsection{Real saddle point: critical point of only the real part of the action}
Finally, we can also consider only the real part of the action to determine the critical point, and perform the integral in a steepest descent approximation about this abridged critical point.  The imaginary part of the action gives a bounded fluctuating contribution, which should be integrable against the measure on the space of functions, provided by the real part of the action.  The real part of the action is just
\beq
f(A)=-\ln ( \cosh\beta m-\cos\beta A).
\eeq
This action is critical at
\beq
f^\prime (A^*)=-{\beta\sin\beta A^*\over \cosh\beta m-\cos\beta A^*}=0
\eeq
which implies that $A^*=0,\pi/\beta, 2\pi/\beta,\cdots$.  The zero at $\pi/\beta$ is a local maximum and thus does not contribute, while the zero at  $A^*=0$ and at $A^{*}=2\pi/\beta$ combine, due to periodicity, to give the full integral over the Gaussian peak.  Then using
\beq
f^{\prime\prime}(A^*=0)={-\beta^2\over\cosh\beta m -1}\label{fpp2}
\eeq
we get
\bea\nonumber
{\cal I}(N,\beta ,m)&=&(\cosh\beta m -1)^N\int_0^{2\pi /\beta }dA e^{-N\ln (\cosh\beta m -\cos\beta A)} e^{iN\beta A}\\
&=&(\cosh\beta m -1)^N\int_{0}^{2\pi /\beta }dAe^{-N\left(\ln(\cosh\beta m -1)+{\beta^2A^2\over2(\cosh\beta m -1)}+\cdots\right)} e^{iN\beta A}.
\eea
This expression defines the perturbation theory about the critical point of just the real part of the action; the $\cdots$ indicates higher order terms that are treated perturbatively.  Now evaluation the integral in the saddle point approximation yields
\bea\nonumber
{\cal I}(N,\beta ,m)&\approx &\int_{-\infty}^{\infty}dAe^{-{N\over 2}\left({\beta^2A^2\over (\cosh\beta m -1)}-i2\beta A\right)} \\
&=&{e^{-{N\over 2}\left(\cosh\beta m -1\right)}\over\beta}\sqrt{2\pi (\cosh\beta m -1)\over N}\cr
\lim_{\beta m\rightarrow\infty}&\to& e^{-Ne^{\beta m}/4}e^{{\beta m/2}}{1\over\beta}\sqrt{\pi \over N}
.\label{ra}
\eea
It is obvious that this result, Eq. (\ref{ra}), is nothing like the result that we obtained by doing the proper steepest descent calculation,  Eq.~(\ref{1}).  To make the contrast  more apparent, we note that in the limit $\beta m\rightarrow\infty$,  the Gaussian approximation controlled by Eq.~(\ref{fpp2}) becomes unsuppressed, {\it i.e.} $\lim_{\beta m\to\infty}{\beta^2\over (\cosh\beta m -1)}\to 0$  while that controlled by Eq.~(\ref{fpp1}), about the critical point of the full action becomes more and more suppressed,  {\it i.e.}   $\lim_{\beta m\to\infty}2\beta^2\coth^2\beta m\to\infty$.  Hence, we conclude, using only the real part of the action to define the perturbation theory, is the wrong procedure to follow.  

\section{Conclusions\label{sec:concl}}
In conclusion, we have shown, via a specific simple example, that the perturbative path integral quantization of field theories with complex critical points requires that we analytically continue into the complex extension of the space of field configurations so that the path of functional integration passes through the complex critical point according to the direction of steepest descent.  Disregarding the complex part of the action to use only the `abridged' critical points, that is, the critical points of only the real part of the action, leads to erroneous results in the perturbation theory.  

\section{Acknowledgements\label{sec:ack}}
We thank Gerald Dunne for useful discussions and the Benasque Center for Science, Benasque, Spain for very pleasant working conditions, where some of this work was done.  We also thank the Inter-University Center for Astronomy and Astrophysics, Pune, India and the Institute of Physics, Bhubaneswar, India for hospitality,  where some of this work was written up.  We thank NSERC of Canada and  the Center for Quantum Spacetime of Sogang University with grant number R11-2005-021   for financial support.

\end{document}